# Ground Based Gravitational Wave Astronomy in the Asian Region

VAISHALI ADYA[1,2], MATTHEW BAILES[1,3], CARL BLAIR[1,4], DAVID BLAIR[1,4], JOHANNES EICHHOLZ[1,2], JORIS VAN HEIJNINGEN[1,4], ERIC HOWELL[1,4], LI JU[1,4], PAUL LASKY[1,5], ANDREW MELATOS[1,6], DAVID OTTAWAY[1,7], CHUNNONG ZHAO[1,4]

[1]AUSTRALIAN RESEARCH COUNCIL CENTRE OF EXCELLENCE FOR GRAVITATIONAL WAVE DISCOVERY OZGRAV
[2]RESEARCH SCHOOL OF PHYSICS, AUSTRALIAN NATIONAL UNIVERSITY, ACTON, AUSTRALIAN CAPITAL TERRITORY 2601, AUSTRALIA
[3]CENTRE FOR ASTROPHYSICS AND SUPERCOMPUTING, SWINBURNE UNIVERSITY, HAWTHORN, VICTORIA 3122, AUSTRALIA
[4]DEPARTMENT OF PHYSICS,, UNIVERSITY OF WESTERN AUSTRALIA, CRAWLEY, WESTERN AUSTRALIA 6009, AUSTRALIA
[5]SCHOOL OF PHYSICS AND ASTRONOMY, MONASH UNIVERSITY, CLAYTON, VICTORIA 3800, AUSTRALIA
[6]SCHOOL OF PHYSICS, UNIVERSITY OF MELBOURNE, PARKVILLE, VICTORIA 3010, AUSTRALIA
[7]DEPARTMENT OF PHYSICS, UNIVERSITY OF ADELAIDE, ADELAIDE, SOUTH AUSTRALIA 5005, AUSTRALIA

*communicated by Leong Chuan Kwek*

**ABSTRACT**

The current gravitational wave detectors have identified a surprising population of heavy stellar mass black holes, and an even larger population of coalescing neutron stars. The first observations have led to many dramatic discoveries and the confirmation of general relativity in very strong gravitational fields. The future of gravitational wave astronomy looks bright, especially if additional detectors with greater sensitivity, broader bandwidth, and better global coverage can be implemented. The first discoveries add impetus to gravitational wave detectors designed to detect in the nHz, mHz and kHz frequency bands. This paper reviews the century-long struggle that led to the recent discoveries, and reports on designs and possibilities for future detectors. The benefits of future detectors in the Asian region are discussed, including analysis of the benefits of a detector located in Australia.

**INTRODUCTION**

Within months of Einstein's publication of the theory of general relativity, Schwarzschild had obtained the solution for spherically symmetric black holes, and Einstein had predicted the existence of gravitational waves (revised and corrected two years later). However, Einstein considered the waves to be of academic interest only, and Schwarzschild's solution was not considered to relate to possible physical reality. There were also ongoing doubts about the physical reality of gravitational waves themselves, by Einstein and others [1].

It was not until the 1957 Chapel Hill conference: *The Role of Gravitation in Physics* that the theoretical reality of gravitational waves was resolved by Feynman, (later elaborated by Bondi and others), with a thought experiment in which gravitational wave strain causes a pair of test masses to do frictional work. In the summary of the conference, Bergmann stated: "there exists a […] type of experiment which is apparently not feasible, and is not going to be feasible for a long time […] the detection of gravitational waves." The conference was attended by John Wheeler who inspired a major theoretical effort in general relativity and black holes, Joseph Weber who pioneered experimental gravitational wave research, and Michael Buckingham who founded gravitational wave research in Australia.

By 1960 Weber had begun an experimental program in gravitational wave detection. In 1962 Freeman Dyson pointed out the immense gravitational wave power emitted if a pair of (then hypothetical) neutron stars were to coalesce [2]. Later, after evidence emerged for black hole powered X-ray binaries, Wheeler and Weber realized that even more power, of order $c^5/G$, would be emitted if black holes were to coalesce. The extra power arises from the fact that black holes are not ripped apart in the final stage of coalescence and reach higher velocities at the point of merger.

Weber built the first resonant mass gravitational wave detectors when the only known potential sources of gravitational waves were supernovae. However, his claims of detection in 1969 and 1970 soon proved to be false, but caused an avalanche of new experiments. The combination of the theoretical possibility of intense gravitational waves and concepts of noise reduction using low loss mechanical resonators explored by Weber motivated an international experimental program in gravitational wave detection that grew steadily from its beginning in the 1970s. Confirming Dyson's prophecy, the first binary neutron star system was discovered in 1976 - the PSR B1913+16 Hulse-Taylor. This provided the first type of gravitational wave source for which precise signals could be estimated.

The experimental program began with laboratory experiments, including cryogenic resonant bars and laser interferometers with length scales of 10 m. Eventually five cryogenic bars across the globe achieved exquisite sensitivity during the 1990s. They were able to rule out frequent gravitational wave bursts occurring in the Milky Way [3]. However, with the hindsight of successful detection, their frequency was too high (700-800Hz), the sensitivity reached an order of magnitude too low, and their bandwidth was too small to detect extragalactic gravitational wave sources.

Kip Thorne, Ron Drever and Rainer Weiss championed the case for building long baseline laser interferometers. This transformed gravitational wave detection from the laboratory to large scale installations. The planning for LIGO in the USA, and Virgo and GEO in Europe began in the 1980s followed by site development and construction in the 1990s. To catch the available technology up to the demanding standards of gravitational wave detection, the projects were arranged to mature over several stages. The initial detectors, as anticipated, did not detect signals, but achieved their science goal of proving key technological advances. On the 14th September, 2015 the two Advanced LIGO interferometers detected a gravitational wave signal from a binary black hole coalescence just after the instruments started collecting data. Since then many more discoveries have enabled the future of gravitational wave astronomy to be mapped out.

In the following sections we review the current status of gravitational wave astronomy. We give an overview of the broader gravitational wave spectrum beyond the reach of current observatories, including proposed space-based detectors and pulsar timing arrays, and followed by a discussion of prominent results of gravitational wave science. We then focus on the future of ground-based detectors and the need for a world-spanning array with increased sensitivity and bandwidth. Finally, we discuss the broad range of state-of-the-art technologies that make these detectors possible, and the frontier areas of research required to realise future detectors.

## THE TECHNOLOGY OF GROUND BASED GRAVITATIONAL WAVE DETECTORS

*The second generation or advanced detectors*

The gravitational wave detector network that successfully detected gravitational waves consists of 3 long baseline Fabry-Perot Michelson interferometers: Two 4 km Advanced LIGO detectors in the USA, and the European 3 km Advanced Virgo detector in Italy. A schematic overview is given in Figure 1. These interferometers convert spatial strain – differential relative length fluctuations of their arms – to phase modulations of their laser car-rier. The current peak sensitivity of ∼$10^{-23}$ 1/√Hz, which requires a displacement sensitivity in the $10^{-19}$ m/√Hz range. This is close to limits set by the quantum uncertainty principles and makes the relative distance measurement the most accurate to date.

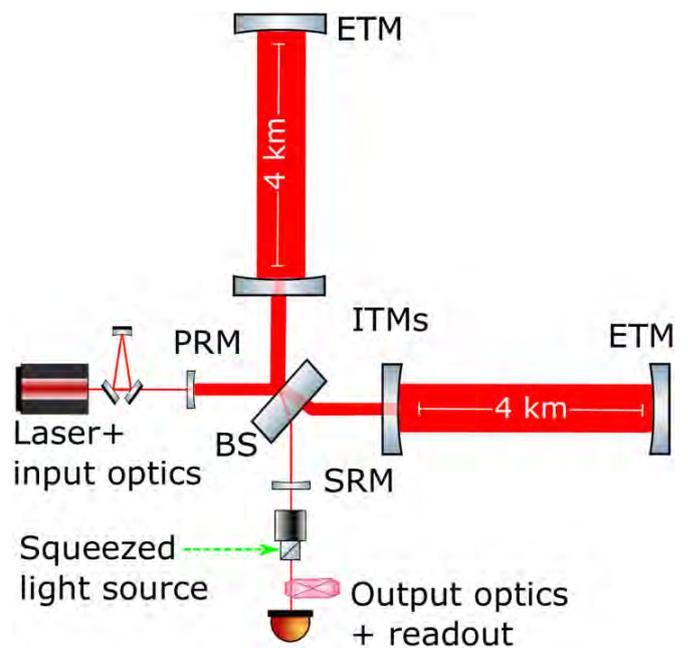

**Fig. 1:** Schematic diagram of Advanced LIGO. The main features are vibration isolated suspended main input and end test mass (ITM/ETM) mirrors and other suspended optical cavity components such as a beam splitter (BS), input and output optics e.g. mode cleaners, and power recycling and signal recycling mirror (PRM/SRM). A low-noise high-power laser and signal detection optics with optical squeezing are used for suppressing photon shot noise.

The performance of detectors is limited by a combination of technical and fundamental noise sources, both classical and quantum in nature. The classical noise sources include seismic noise, which has prompted the development of superior suspension systems [4, 5], and thermal noise due to dissipation in the mirror substrates and coatings [6, 7], a consequence of the fluctuation-dissipation theorem [8]. The quantum noise manifests itself as radiation pressure back-reaction on the mirrors or *test masses* at low frequencies and photon shot noise at high frequencies [9, 10]. In addition, a vast array of technical noise sources can contribute noise through light scattering, control systems, residual gas, unwanted optical modes, mirror thermal distortions, point defects in optical coatings and other processes.

To overcome the sources of fundamental noise, the detectors combine multiple advanced technologies including high power, high stability 1064 nm lasers, multi-stage monolithic suspensions, and 40 kg fused silica test mass mirrors. The mirrors have exceptionally high homogeneity and purity combined with superpolished optical surfaces that are precise to the sub-nanometer level. Optical coatings on the mirror surfaces have extremely low optical losses and low acoustic losses.

The high frequency sensitivity is largely determined by shot noise on the readout photodetector due to the Poissonian distribution of photon arrival times. Higher laser power results in more photons counted, which decreases the associated statistical uncertainty with the square root of the optical power in the arms. For this reason, 800kW of circulating optical power in the two Fabry-Perot cavities were initially proposed for Advanced LIGO This has been difficult to achieve mainly due to thermal distortions of the mirrors that arise from point absorbers in the coatings. The strongly localised absorption of laser light spot-heats the mirror, warping their surfaces and creating inhomogeneous dispersion profiles. As an additional measure to lower quantum noise, a quantum-squeezed vacuum state, void of photons but with reduced zero-point fluctuations in a preferred quadrature, is injected into the output port. The squeezed state is superimposed with the laser output and reduces the quantum fluctuations, which we observe as shot noise, in the readout.

*Upcoming detectors: KAGRA, LIGO-India and A+*
Current plans are for aLIGO and Virgo to reach design sensitivity by ∼2021, with a 2.5-generation detector A+ due to reach design sensitivity circa 2024 [11]. Other second-generation instruments currently being commissioned and built are the Japanese KAGRA detector and LIGO India, due to be online in late 2019 and the mid 2020s, respectively [11]. KAGRA is pioneering two pivotal approaches to gravitational wave detection: it is the first underground detector, sharing the infrastructure of the well-known Kamioka mine to house its 3 km arms, and is also the first detector to use cryogenics as a measure to reduce thermal noise. Its in-vacuum suspended sapphire test masses need to be cooled to 20 K, which presents significant technical challenges, and the generated experience will define the roadmap for proposed future detectors. KAGRA is projected to join coordinated observation runs with LIGO and Virgo at the end of 2019.

LIGO-India will be a technological clone of the LIGO detectors, with 4 km arms. Starting in the early-2020's LIGO will be upgraded to LIGO A+. It will use improved mirror coatings designed to reduce both the thermal noise of the mirror coatings and point defects. Soon afterwards, LIGO-India will be commissioned using identical technology. Currently, the site for LIGO-India has been acquired near Aundha in the state of Maharastra in west India. A design for the infrastructure is in an advance state. See Figure 2 for the proposed sensitivity curve of all three LIGO A+ facilities.

To increase the volume of the Universe accessible to gravitational wave astronomy, the detector network alternates extended observation periods with commissioning breaks for the individual detectors to perform maintenance tasks and tweak operational parameters. Major upgrades, which are more intrusive to the infrastructure but promise significant improvements with the integration of new technologies, are generally spaced by several years.

Advanced LIGO will commence upgrade activities to the 'A+' stage of the project in mid-2020. Cornerstones of the upgrade are new test masses with novel reflective coatings that are less dissipative and have better optical quality, and the extension of the vacuum envelope to include a filter cavity for the back-port injected squeezed light. Much research effort has gone into reducing the mechanical loss in the tantala-silica dielectric coatings without spoiling optical properties using dopants, varied deposition techniques, and optimised annealing strategies. The filter cavity imposes a frequency-dependent phase on the injected squeezed vacuum field. When controlled correctly, this results in amplitude squeezing at low frequencies and phase squeezing at high frequencies, for a simultaneous improvement in shot noise and radiation pressure noise.

**Proposed Detectors**

Three new terrestrial detectors are proposed. Two of them are described as third generation or 3G detectors, which aim for an order of magnitude improvement in sensitivity over the 2G or advanced detector design sensitivity. The other is a high frequency detector.

*Third generation (3G) detectors*

In Europe, the Einstein Telescope (ET) observatory [12] is proposed to be a 10km underground equilateral triangle containing six detectors – both a low frequency cryogenic detector (ET-LF) and high frequency high power (ET-HF) detector at each of three corner stations. In the USA a detector called Cosmic Explorer (CE) [13] has been proposed, that would have 40km arms. Both would achieve comparable sensitivity. The increase in arm length is required for suppressing noise generated locally at the test masses. The gravitational wave strain is defined as h = ΔL/L, where L is the arm length and ΔL is the induced change in arm length. If our test masses have intrinsic noise ΔL, then the only means of improvement is by extending the arm length. Test mass noise acts most strongly at around 100 Hz. At this frequency the only choices are to either to reduce the loss of the test mass coatings, or to increase the arm length. Increased arm length is not sufficient for increasing the broad-band sensitivity. Both of the above designs propose increasing the optical power to 3 MW (ET-HF) and 2 MW (CE).

At frequencies above 1 kHz, thermal noise is not limiting and as discussed in the previous section there are extremely interesting signals in this band. For his reason, there is great interest in the concept of building high frequency detectors, that would not have to use very long (and expensive) arm lengths, and which could achieve sensitivity at 1-5 kHz comparable or better than that proposed for the 3G detectors. One of these concepts is the OzHF concept discussed below.

*High frequency detectors*

The goal of high frequency detectors is to achieve sensitivity around $10^{-24}$ Hz$^{-1/2}$ between 1-5 kHz. Most designs use a configuration described as strongly coupled, detuned signal recycling. The basic configuration is similar to Advanced LIGO. However, the signal recycling cavity, which resonantly amplifies the gravitational wave signal sidebands is tuned to a frequency several kHz higher than the main carrier frequency. When this is combined with an increase in circulating optical power, to 5 MW, the sensitivity goal of $10^{-24}$ Hz$^{-1/2}$ can be achieved in a 4 km length interferometer. New technologies discussed below may be able to accommodate the very high power.

Given the technical risk of a design that requires very high optical power, there are two ways of reducing this power requirement. The first recognizes there is some trade-off between arm length and optical power. Increasing the arm length is relatively risk free, but expensive. Any proposed detector must optimize both the costs and the risks. The latest sensitivity curve for a 4 km design, where the signal recycling cavity length and reflectivity is optimised for that armlength, is shown in Figure 2.

The second approach is an optomechanical technique called white light signal recycling which gain sensitivity without requiring very high optical power. It involves using an optomechanical filter in the interferometer output path that compensates the dispersion of the signal sidebands in the recycling cavity. If correctly compensated, the cavity can combine high resonant gain with high bandwidth. Much research is underway attempting to demonstrate this technique.

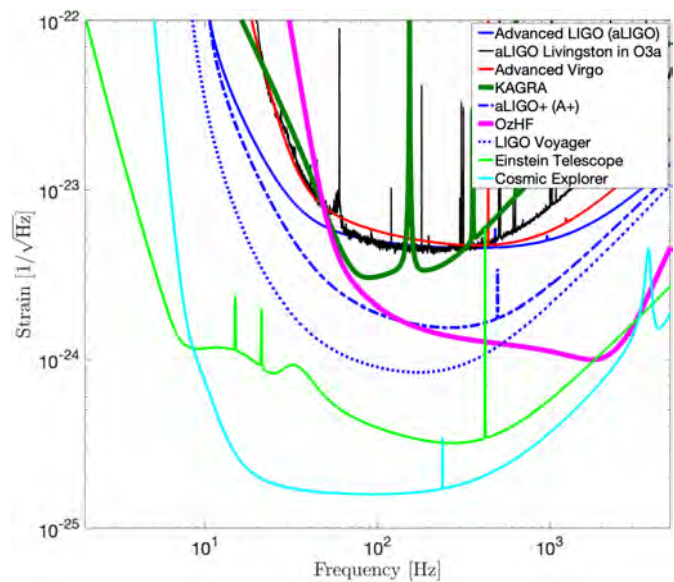

**Fig. 2:** Sensitivity curves for current, future and possible gravitational wave detectors. The sensitivity obtained for LIGO Livingston at the beginning of Observation Run 3 (O3), which runs from April 2019 – April 2020, is plotted. The OzHF trace is the most up-to-date curve of this proposed high-frequency detector. More details are found in the text.

## THE GRAVITATIONAL WAVE SPECTRUM

Although gravitational waves have now been detected in ground-based detectors in the 10s to 100s of Hz range, there are exciting prospects for their detection in other bands in the coming decades [14]. Millisecond pulsars are naturally-occurring celestial clocks that are rapidly-rotating neutron stars that emit a coherent beam of radio emission. Fortunately, the galaxy possesses some 100,000 active millisecond pulsars and these form a galaxy-sized gravitational wave detector that is sensitive to supermassive black hole binaries in the nanohertz regime. Radio astronomers have been discovering these objects since the early 1980s and by now their number has grown to almost 300, nearly 100 of which are suitable for monitoring in the International Pulsar Timing Array. State of the art detection yields timing precision at the 100-nanosecond level [15], and correlations of timing between pulsars that are close in the celestial sphere will be a tell-tale sign of the presence of a gravitational wave background. New facilities such as the FAST telescope in China [16], the MeerKAT telescope in South Africa [17] and upgraded facilities like the Parkes 64m Ultra-wide band receiver are rapidly advancing the race for the detection of nanohertz gravitational waves.

In the millihertz regime, the LISA detector [18] is planned for launch in 2034. The LISA mission will fly three spacecraft in triangular constellation, separated by millions of kilometres, and will detect gravitational waves from a multitude of sources. The signals will probably be dominated by the galactic foreground of close white dwarf binaries. Frequently, however, stronger sources will rise above the background and give witness to the coalescence of large mass black holes in the more distant universe. LISA will also detect binary neutron star systems well in advance of the ground-based detectors, and even predict their eventual merger at ground-based frequencies, like the kilonova that accompanied GW180817. There are also similar Chinese space gravitational wave detector projects, TianQin [19] and Taiji [20], which are planned to launch in the 2030s.

One as-yet-unobserved prediction of inflation is that it would produce gravitational waves, which can be observed through their signature on the cosmic microwave background polarization. Models of inflation predict that gravitational waves will source B-modes at angular scales of a degree or larger. In March 2014, a discovery of just such a signal with BICEP2 was reported, but subsequent results from the Planck satellite have shown that Galactic foregrounds, specifically dust, accounts for most if not all of the B-mode signal. The BICEP collaboration and other experiments have since increased efforts to improve their measurements.

## THE TIP OF THE ICEBERG: RECENT RESULTS AND THE FUTURE OF GRAVITATIONAL WAVE ASTRONOMY

The era of advanced gravitational-wave detectors has surpassed most expectations. The first two LIGO/Virgo observing runs O1 and O2, respectively, provided the first ten gravitational-wave signals from binary black hole mergers [21]. The science output from these detections is remarkable. Among other things, this includes probing gravity in the ultra-strong field regime, understanding the formation history of these exotic systems and beginning to inform our understanding of supernovae, for example the role pair instability supernovae play in cosmic history [e.g., 22, 23].

The second observing run O2 also provided the first gravitational-wave observation of a system of coalescing binary neutron stars, GW170817 [24]. The gravitational-wave observations alone provide some of the tightest constraints to date on the equation of state of nuclear matter at supranuclear densities through measurements of the neutron stars' tidal deformability [24]. Prompt observations by the Fermi-GBM and INTEGRAL gamma-ray burst detectors firmly established the association between short duration gamma-ray bursts and binary neutron star mergers [25, 26, 27]. Furthermore, the sky localization to within 28 square degrees provoked a ground-breaking electromagnetic follow-up campaign that initialised the era of gravitational wave multi-messenger astronomy [25].

The electromagnetic follow-up campaign of GW170817 delivered on many fronts. The optical, ultraviolet and infrared observation of a kilonova—the glow produced by the radioactive decay of the neutron-rich ejector—evolved from blue to red over a week [28]. Late-time observations from the X ray to radio bands carefully monitored the afterglow before eventually showing that a successful jet was launched and that the first emissions were from a structured jet viewed approximately 20 degrees from the jet axis [29]. Again, the scientific output was immense. For example, the 1.74-second delay between the gravitational-wave inferred merger time and the gamma-ray observations placed the tightest constraints on the speed of gravity by many orders of magnitude [25]. The kilonova observations gave insight into the nucleosynthesis of heavy elements [30]. The colour of the kilonovae indicated a nascent neutron star survived the initial merger but collapsed into a black hole which, if correct, places some of the most stringent constraints on the maximum neutron star mass [e.g., 31] The gravitational-wave and galactic localisation provided the first estimate of the Hubble constant in the local Universe, independent of the cosmic distance ladder [32]. This measurement was further refined with late-time radio observations giving insight into the inclination angle of the binary, and therefore placing even tighter constraints on the Hubble parameter [33].

The third observing run O3 is broken into two, with O3a going from April 1 to September 30, 2019 and O3b starting on November 1 to end April 30, 2020. Like previous observational runs, this has exceeded expectations. During O3a, public GCN notices show that approximately one merging black-hole candidate has been observed every week along with a handful of coalescing binary neutron star and black-hole neutron star candidates [34]. Detailed analysis of the gravitational-wave signals from these events is ongoing. The increased number of detectors across the globe significantly increases our ability to localise gravitational-wave sources, and hence perform electromagnetic follow-up of gravitational-wave events.

Figure 3 shows the expected annual detection rates for binary neutron star mergers as a function of redshift for aLIGO at design sensitivity, A+, and the third-generation detectors currently slated for the 2030s. While we expect aLIGO to detect 10s of events per year, the A+ upgrade will detect 100s of binary neutron star mergers out to cosmological distances. Cosmic Explorer, on the other hand, will routinely detect such events out to redshifts of two, and binary black hole mergers out to redshifts of 20 [35].

The accumulated catalogue of binary black hole mergers in the third-generation era will allow unprecedented insights into the formation of these exotic system through an understanding of their masses, spins, and eccentricity distributions, both as a snapshot from the local Universe and as a function of cosmic time. Moreover, the accumulation of events during both the Advanced LIGO, A+, and Cosmic Explorer/Einstein telescope era will allow data from multiple binary black hole collisions events to be coherently combined to provide insights into the nature of gravity in the strong-field regime, for example by testing the no-hair theorem of general relativity, testing the black-hole area theorem, and measuring gravitational-wave memory.

The increased sensitivity of gravitational-wave detectors, and the introduction of large field-of-view telescopes such as the Large Synoptic Survey Telescope (LSST), among many others, will significantly enhance the number of coincident gravitational-wave and electromagnetic observations of binary neutron star mergers. These multi-messenger observations will provide valuable insights into the inner engines of these high-energy transients, including into the physics of their relativistic jets and subsequent interactions with the interstellar medium. Third-generation gravitational-wave detectors have the potential to detect the inspiral of such events, and provide negative-latency triggers that alert telescopes to the merger *before* it has taken place. Such electromagnetic follow-up campaigns will be aided by the increased angular resolution offered by future ground- and space-based telescope networks.

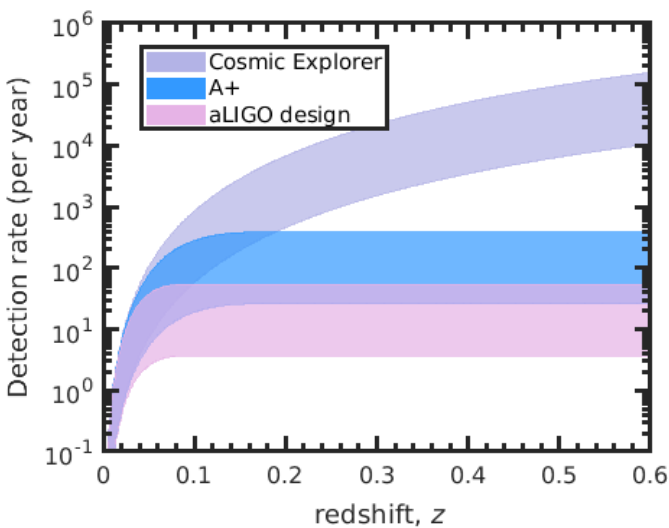

**Fig. 3:** The detection rate of coalescing binary neutron star systems for different detection epochs. The colored bands cover the range of rate estimates for each epoch. At advanced LIGO design sensitivity (2021) tens of events will detected increasing to hundreds of events each year by the time of A+ (2024); Cosmic Explorer (2030s) will essentially detect every event in the visible Universe.

Table 1 shows how the network angular resolution improves every time a new detector is added to the worldwide array. Notably, we see that an Australian detector improves the pointing accuracy in all cases. For illustration we assume here a binary black hole coalescence like GW150714 at 400Mpc. This improvement can be critical in resolving source host galaxies for cosmological studies and pinpointing electromagnetic counterparts for rapid follow-up campaigns. Furthermore, as the network increases so too does the live time or duty cycle increase the number of detections. A larger network also improves detection sensitivity allowing greater precision on the spins and polarizations of the sources.

**Table 1:** Network angular-resolution improvement every time when new detectors are added to the global array of gravitational-wave detectors. The table compares the existing network of the two aLIGO detectors (HL) and the AdVirgo detector (V) with future network configurations including detectors in Japan (J) and India (I) and an Australian based detector (A). The percentages shown assume a source equivalent to the first gravitational wave detection GW150914 at 400 Mpc.

| Network | Percentage of detections within skymap area | | | |
|---|---|---|---|---|
| | 0.1 deg$^2$ | 0.5 deg$^2$ | 1 deg$^2$ | 5 deg$^2$ |
| LHV | 0% | 1% | 8% | 45% |
| LHVJI | 0% | 27% | 52% | 91% |
| LHVJIA | 7% | 67% | 88% | 100% |

*Peering into the unknown*

To date, LIGO and Virgo data has only shown evidence of individually-resolved compact binary coalescences. But the future of the field is rich, replete with known and expected sources, those that are classified as speculative at best, and everything in between. Here we review some of these potential new sources that may be unveiled with the current and future generations of gravitational-wave detectors.

Immediately following the merger of two neutron stars, there are two main classes of outcome. First, the object collapses promptly to form a black hole, in which case the subsequent gravitational-wave emission can only be seen if the source is within our Galaxy. On the other hand, a nascent neutron star may be born rapidly rotating and with an ultra-strong magnetic field. Such systems are believed to emit gravitational waves in the kHz regime, potentially detectable out to GW170817 distances with third-generation technology [e.g. 36 and references therein]. The primary gravitational-wave emission modes of such an object trace the equation of state of nuclear matter at non-negligible temperatures (unlike during the binary inspiral phase) [37]. These post-merger remnants therefore provide a new cosmic for studying quantum chromodynamics in a regime not accessible elsewhere in the Universe. Designs for dedicated high-frequency detectors, with the primary science goal of detecting such post-merger remnants, are in development (see below).

Binary neutron star systems collide somewhere in the Universe approximately every 13 seconds, and binary black holes collide approximately every 5 minutes [38]. However, most of these are too distant to be detected. Such sub-threshold signals form a stochastic gravitational-wave background that can be searched for in ground-based interferometers using varying techniques. Astrophysical stochastic backgrounds are formed by such compact binary coalescences, and also supernovae, isolated neutron stars, and more exotic phenomena such as cosmic strings, while cosmological backgrounds may also be present as relics from inflation or phase transitions in the early Universe. Current estimates suggest the stochastic background from coalescing black holes and neutron stars may be detectable with current sensitivity [e.g., 39, 40].

The first detection of gravitational waves from a core-collapse supernova would be a discovery of enormous historical significance with significant scientific implications. What ignites the explosion in a dying star is one of the outstanding questions in astronomy. Gravitational waves that travel unimpeded by gas and dust could provide insights on the internal mechanisms that produce these spectacular explosions. In contrast to coalescing systems, the gravitational-wave emission from core-collapse supernovae are expected to be relatively weak; current estimates suggest the second-generation network is only sensitive to supernovae within our galaxy [41]. Although core-collapse supernovae explode somewhere in the Universe around 100 times each second, only 1-2 occur in our galaxy every century. Such detections may therefore have to wait for third-generation gravitational-wave detectors that could possibly detect these explosions out to around 10s of Mpc.

In addition to supernovae, gravitational-wave searches of LIGO/Virgo data are also conducted for other transient phenomena such as pulsar glitches, cosmic strings, or the physical mechanism behind fast radio bursts [42]. Such searches are often referred to as un-modelled, in the sense that the gravitational waveform is largely unknown.

Not only are the signal amplitudes of these expected signals unknown, but in many cases so is the event rate. Speculation on when the first detection of these systems therefore abounds.

In addition to transient events, nearly monochromatic signals are expected, most likely from asymmetries on rotating neutron stars [e.g., 43]. From a theoretical perspective, the unknown quantity is how large mountains or oscillation-mode amplitudes (such as inertial r-modes) on neutron stars can be, and therefore how large is the subsequent gravitational-wave amplitude. Non-detections of nearly-monochromatic gravitational waves from known systems [44] and all-sky searches [45] provide constraints on the maximum size of these mountains and oscillation amplitudes. Radio observations may place a lower limit on the size of mountains on millisecond pulsars, which in turn suggests detection of such a signal could be possible with aLIGO/Virgo operating at design sensitivity [46]. However, that work is speculative at best, and the required gravitational-wave sensitivity and time-to-detection for a nearly-monochromatic gravitational-wave signal from a neutron star is largely unknown.

## NOVEL TECHNOLOGY FOR FUTURE DETECTORS

*Squeezing and frequency dependent optical squeezing*
Vacuum fluctuations that enter the interferometer through the open/unused ports cause quantum noise. In order to improve the broadband sensitivity of the detector, one needs to counter quantum radiation pressure at low frequencies and quantum shot noise at high frequencies [10, 47]. Nonlinear optics are used to squeeze the vacuum that leaks into the empty port of the beam splitter. This essentially means second order nonlinear PPKTP crystal to split 532 nm light into two correlated 1064 nm photons [48].

The amount of squeezing is generally expressed in decibels, normalised to the uncertainty of the un-squeezed vacuum state. Currently, the LIGO detectors achieve 2.7 dB of squeezing at Livingston and 2.0 dB at Hanford. This improves the range of the detectors to binary neutron star mergers by 15% and 12% respectively. The detection rate, which scales with the volume of the observable Universe, consequently increases by 50% and 40% for the two detectors, respectively compared to no squeezing [49]. Future detectors aim for higher levels of squeezing by improving the lossy elements in the squeezing injection path, such as the Faraday isolator, and active spatial mode-matching between the squeezed field and the interferometer.

The relationship between the entangled photons produced by the squeezer can shaped using a filter cavity to suppress phase noise at high frequencies and amplitude at low frequencies because unfortunately it is not possible to do both at the same frequency when just injecting squeezed vacuum states. This results in optimal squeezing angle being different at low frequencies and high frequencies. This type of squeezing – called frequency dependent squeezing - solves this problem by filtering the injected squeezed field by reflecting it off an optical cavity with a similar linewidth to the cavity seen at the dark port of the interferometer and thereby creating the required phase rotation.

The implementation for the A+ LIGO detectors will use a 300 m cavity to meet the linewidth and loss requirements for the frequency dependent squeezing cavity [50]. It is possible to use an opto-mechanical cavity just 12 cm long to meet the linewidth requirements [51]. The optical loss requirements of the conventional cavity in this case are conflated with the phase noise introduced by the mechanical resonator and therefore the requirements of the optomechanical white light cavity provide a stepping stone to the development of an optomechanical filter cavity for squeeze angle rotation. All these improvements would allow A+ to obtain its design sensitivity with 6 dB reduction in quantum noise. Future detectors designs have similar squeezing designs to improve sensitivity up to 10 dB for ET and CE.

*2-micron cooled silicon optics*
The advantage of switching to crystalline silicon substrates is two-fold. First, silicon has very low intrinsic mechanical loss, which further improves when cooled [52]. This is in contrast to fused silica, which has a loss peak at 20 K that is four orders of magnitude higher than its room temperature value. Second, silicon also has far better thermal conductivity, which significantly reduces the temperature gradients in the substrate when spot-heated by a high-power beam. This will greatly help to reduce thermal lensing, an ongoing issue in the current detectors, which limits the optical power in the recycling and arm cavities. High-purity silicon boules grown by the magnetic Czochralski method are projected to become available in diameters up to 45 cm over the next decade, an important prerequisite for developing test masses.

The 1 μm laser wavelength used in current detectors lies close to the band gap in silicon, and would be too strongly absorbed in the new test masses. It will be replaced by a longer wavelength in the 1.5 μm to 2 μm region.

One caveat is higher thermo-elastic noise in silicon, which is driven by random temperature flows in the substrate, another consequence of the fluctuation-dissipation theorem [53]. Thermal expansion couples the resulting local temperature fluctuations to test mass deformations, distorting the wavefronts and generating displacement noise. Conveniently, silicon's thermal expansion coefficient crosses zero at 123 K and 18 K. Operating at these temperatures strongly suppresses thermo-elastic noise, allowing cryogenic improvements of coating Brownian noise to fully unfold, and further mitigates thermal lensing.

The current tantala-silica-based dielectric coatings are not suitable for cryogenic loss improvements of Brownian noise. Tantala is the culprit for their high compound mechanical loss at room temperature, and there is no significant relief from cryogenic cooling [54]. For Advanced LIGO, doping the tantala layers with titania and optimising the post-deposition annealing treatment has provided incremental loss improvements, which is further pursued for the A+ upgrade [55]. For the order of magnitude improvement targeted by future detector proposals, better coating materials need to be identified.

Low optical absorption is an additional requirement for any candidate coating to keep the heat load on the test masses minimal. One of the more promising high index materials compatible with the 1.5 μm to 2 μm wavelength region is amorphous silicon (α-Si). Like thin-film silica, its loss angle is an order of magnitude better than tantala at room temperature, and it further improves with cryogenic cooling [56]. The high refractive index of 3.5 would also reduce the number of coating layers required for given reflectivities, as well as the individual layer thicknesses. Amorphous silicon does show a small amount of excess absorption, but it is likely that this can be reduced with deposition process control and optimised annealing strategies [57].

A different approach is the use of epitaxially grown crystalline thin-film structures as optical coatings. Extremely low mechanical loss across a broad range of temperatures, as well as low absorption and optical scatter have been demonstrated in GaAs-AlGaAs Bragg mirrors [58]. The biggest hurdle for this technology is the cost involved in scaling it up to the mirror size proposed for next generation detectors, which requires large area Gallium-Arsenide wafers.

*Optomechanical white light cavities*
The sensitivity bandwidth of current gravitational wave detectors is limited by the linewidth of the optical cavity seen at the output of the interferometer. This cavity is composed of the signal recycling cavity coupled to the arm cavities of the interferometer, or just the arm cavities in the case that there is no signal recycling cavity.

The bandwidth of the detector can be improved if the linewidth is increased, however conventional approaches trade bandwidth for peak sensitivity. In principal it is possible to increase detector bandwidth without reducing peak sensitivity by introducing a negative dispersion medium into the optical cavity. Generally speaking, the sensitivity of the gravitational wave detectors presented in figure 3 can be extended with a flat region in the sensitivity curve from 1-4 kHz – rather than the gradual rise - by use of the white light cavity technique.

The negative dispersion medium advances the phase of lower frequencies, while retarding phase of higher frequencies resulting in a broader frequency range satisfying the resonance condition of the cavity. An optomechanical cavity operating in the unstable blue detuned regime can provide negative dispersion assuming additional feedback control for stabilisation. The challenge in gravitational wave detectors is in creating a negative dispersion medium that satisfies strict requirements on optical loss and phase noise imparted to the optical field. Equally demonstration of the technique in gravitational wave detector models will be an important part of demonstrating feasibility. Negative dispersion has been demonstrated in an opto-mechanical cavity with a tunable linewidth that could compensate the round-trip phase dispersion - which is positive - of 4 km arm of Advanced LIGO for a broadband range of frequencies. [59]

Opto-mechanical cavities can in principle meet the strict requirements for optical loss and phase noise to broaden the linewidth of the signal recycling cavity in a detector like Advanced LIGO [60]. However, the phase noise requirement results in a requirement that the thermal noise of the mechanical system satisfy $Q/T > 10^{10}$, where Q is the effective quality factor and T is the temperature of the mechanical oscillator.

Two approaches are being pursued to meet this strict requirement. First optical dilution is being tested on microresonators presented in the above reference. Optical dilution is a method where radiation pressure from an optical field is used to create an optical spring to compliment the mechanical spring. As the optical field has low dissipation, the resulting optomechanical system can have an enhanced effective Q factor. Second, incredibly low loss resonators used in the telecommunications industry [61] are being tested.

## DISCUSSION AND CONCLUSION

The global array of gravitational wave detectors will increase to five when KAGRA and LIGO-India come on line. As emphasised above the network angular resolution, sensitivity and duty cycle improve every time a new detector is added to the array. This has been well demonstrated by events seen recently by two or three of the LIGO and Virgo detectors. However angular resolution depends on the spacing between detectors. A new detector located in Australia creates multiple new baselines that are close to the maximal possible 40 ms light travel time spacing possible on the Earth's surface. Better angular resolution allows the host galaxies of sources to be identified, which greatly facilitates cosmological studies.

*A Possible Asian-Australian detector*
The high frequency detector concept discussed above provides a possible starting point for a southern hemisphere detector. It could begin as a relatively low-cost detector located at a site in Australia where it would be practical to plan future extension of the arm length. It could begin by focusing on neutron star coalescence observations, but be extended and improved to encompass the full bandwidth available for ground-based detectors.

The proposed detector should be an international project. It would provide an opportunity for physicists across the east-Asian time zone range to collaborate in a project too large for individual countries. The detector would make an enormous contribution to the opening of the gravitational wave spectrum. It would be well matched to many advanced technologies associated with semiconductor industries across the region, and would be a powerful vehicle for positive cooperation and cultural exchange in the region.

Gravitational wave astronomy has linked scientists across the world in making dramatic new discoveries of the universe. The detectors link a broad range of research areas: mechanical engineering, quantum optics and quantum measurement, materials science, semiconductors and coatings, laser physics, geophysics, and data science. The discoveries have linked gravitational wave astronomers with gamma and X-ray astronomers, radio astronomers and optical astronomers. A future detector could link the diverse countries in the east Asian time zone in a common endeavour that would enhance science, education and international relations across the region.

**Acknowledgements:** This work was funded by the Australian Research Council (ARC) Centre of Excellence for Gravitational Wave Discovery OzGrav under grant CE170100004. EH is supported through ARC DECRA Fellowship DE170100891. PL is supported through ARC Future Fellowship FT160100112 and ARC Discovery Project DP180103155. CB is supported through ARC DECRA Fellowship DE190100437.